\documentclass[twocolumn,showpacs,preprintnumbers,amsmath,amssymb]{revtex4}
\usepackage{graphicx}% Include figure files
\usepackage{dcolumn}% Align table columns on decimal point
\usepackage{bm}% bold math

\input{psfig.sty}
\newcommand{\be}{\begin{equation}}
\newcommand{\bea}{\begin{eqnarray}}
\newcommand{\ee}{\end{equation}}
\newcommand{\eea}{\end{eqnarray}}

\begin{document}
\preprint{}
\title{ Synaptic plasticity of Inhibitory synapse promote synchrony in inhibitory network in presence of
heterogeneity and noise}% Force line breaks with \\

\author{Sachin S. Talathi}%
 \email{talathi@physics.ucsd.edu}
\affiliation{%
Department of Physics and Institute for Nonlinear Science \\
University of California, San Diego, La Jolla, CA 92093-0402, USA
}
\date{\today}
\begin{abstract}
Recently spike timing dependent plasticity was observed in inhibitory synapse in the layer II of entorhinal cortex. The rule provides an interesting zero in the region of $\Delta t=t_{post}-t_{pre}=0$ and in addition the dynamic range of the rule lie in gamma frequency band. We propose a robust mechanism based on this observed synaptic plasticity rule for inhibitory synapses for two mutually coupled
interneurons to phase lock in synchrony in the presence of intrisic heterogeneity in firing. We study the stability of the phase locked solution by defining a map for spike times dependent on the phase response curve for  the coupled neurons. Finally we present results on robustness of synchronization in the presence of noise.
\end{abstract}
\pacs{Valid PACS appear here}
\maketitle

It is generally accepted that inhibitory interneurons are
important for synchrony in neocortex. Several studies have
reported the role for inhibitory interneurons in generating stable
synchronous rhythms in neocortex
~\cite{Benardo,Jefferys,Michelson,Whittington,Bragin}. Cortical
oscillations in the gamma frequency band (20-80 Hz), are thought
to be involved in binding of object properties, a process of great
significance for the functioning of the brain. These experimental
findings have led to numerous theoretical studies of synchrony
among inhibitory interneurons ~\cite{van, Ernst, Traub, Wang}. The
principle result of these studies showed that depending on the
decay time of the inhibitory synaptic coupling, the mutually
coupled inhibitory neurons oscillate in synchrony ( in phase
locking) or in antisynchrony (out of phase locking). However much
of the above investigations did not explore the effects of
heterogeneity in the intrinsic firing rates nor did they take into
account noise, which is invariably present in neuronal systems.

In another set of theoretical investigations, ~\cite{White}
explored the implications of small heterogeneity for the
degradation of synchrony of fast spiking inhibitory neurons and
the mechanism by which the degradation occur. They found that
introduction of even small amounts of heterogeneity in the
external drive, resulted in significant reduction in coherence of
neuronal spiking. It is important then to understand what mediates
observed in vivo synchrony of neuronal networks under biological
realistic conditions of noise induced unreliability and intrinsic
heterogeneity in spiking rates of the neuronal ensemble.

\begin{figure*}[ht!]
\includegraphics[width=5.0in,scale=1,angle=0]{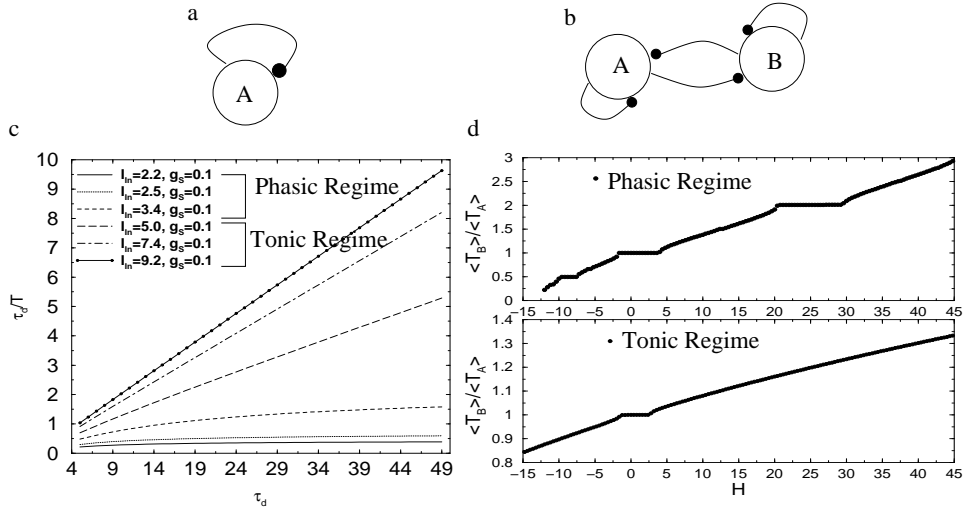}
\caption{(a) Schematic diagram of the self inhibited neuron
considered to determine the different regimes of operation of
network of mutually coupled interneurons. (b) Schematic of the
mutually coupled interneurons. (c) Ratio of the synaptic decay
constant $\tau_{d}$ to firing period T plotted versus $\tau_{d}$.
Parameter values for the operation of the neuron in the two
regimes, are shown. (d) The ration of firing period of two coupled
neurons in phasic and tonic regimes are plotted as function of
heterogeneity in external drive.\label{Fig1}}
\end{figure*}

In this letter we propose a robust mechanism based on spike timing dependent synaptic plasticity of
inhibitory synapses ~\cite{Haas}
 by which two coupled interneurons can phase lock in synchrony
even under conditions of mild heterogeneity in the firing rates of
the coupled neurons and in the presence of noise. Earlier work
\cite{Valentin} has explored the function of synaptic plasticity
in the excitatory synapse in improving synchronization in
unidirectionally coupled neuronal network. We consider a network
of two coupled interneurons with self inhibition as shown in
Figure \ref{Fig1}b. The self-inhibition is introduced because
biological neural networks often have
 local inhibitory interneurons which deliver feedback inhibition to the cells activating those
 interneurons ~\cite{Sheperd}. The importance of self inhibition that simulates the network effect, is explored in
 details in \cite{Talathi_Neurosci}.

 Each neuron in the coupled network is modelled as
\begin{eqnarray}
\frac{dV_{i}(t)}{dt}&=&I^{i}_{In}+g_{Na}m^{3}(t)h(t)(E_{Na}-V_{i}(t)) \nonumber \\
&+&g_{K}n^{4}(t)(E_K-V_{i}(t))+g_{L}(E_L-V_{i}(t))  \nonumber \\
&+&I^{ij}_{M}(t)+I^{i}_{S}(t) +\eta \zeta_{i}(t) \nonumber \\
\end{eqnarray}
where $V_{i}(t)$ (i=A,B) is the membrane potential ,
$\zeta_{i}(t)$ is the gaussian synaptic noise of amplitude $\eta$ satisfying $\left<\zeta(t)\right>=0$ and
$\left<\zeta_{i}(t)\zeta_{j}(t^{'})\right>=\delta(t-t^{'})\delta_{ij}$.
$I^{i}_{In}$, is the external drive, $I^{i}_{S}(t)=g_{s}S(t,V_{i}(t))(E_{I}-V_{i}(t))$, is the synaptic current due to self inhibition and
$I^{ij}_{M}(t)=g_{j\to i}S(t,V_{j}(t))(E_{I}-V_{i}(t))$ is the synaptic current from mutual
inhibition. $g_{j\to i}=g_{m}G(t)$ is the dynamic synapse, whose strength is determined by the
inhibitory synaptic plasticity rule, and $g_{s}$ is the synaptic strength of self inhibition.
$E_r$ (r=Na, K, L) are reversal potentials of the sodium and
potassium ion channels and the leak channel respectively. $E_{I}$ is the reversal potential of
the inhibitory synapse. $S(t,V(t))$ give the fraction of bound receptors and satisfy the first
 order kinetic equation, $\dot{S}(t)=\frac{S_{0}(V_{pre}(t))-S(t)}{\hat{\tau}(S_{I}-S_{0}(V_{pre}))}$
 where $V_{pre}(t)$ is the presynaptic voltage. It involves two time constants, $\tau_{r}=\hat{\tau}(S_{I}-1)$, the docking time for
the neurotransmitter and $\tau_{d}=\hat{\tau}S_{I}$, the undocking time constant for the neurotransmitter binding.
 $S_{0}(V)$ is the sigmoidal function given by,
 $S_{0}(V)=0.5(1+\tanh(120(V-0.1)))$. The gating
variables X(t),(X=m,h,n), satisfy first order kinetic equations,
 $\dot{X}(t)=\alpha_{X}(V)(1-X(t))-\beta_{X}(V)X(t)$.
We have used standard functions $\alpha_{X}(V)$ and $\beta_{X}(V)$ and parameters for the model \cite{Talathi},
 such that the dynamics of the neurons to spiking is through saddle node bifurcation and
 the model represents a type I neuron \cite{Ermentrout}.
 The model parameters are within physiological range and give high spike rates typical of interneurons.

We consider two regimes of operation of the network shown in Figure \ref{Fig1}b. The two regimes
called the phasic regime and the tonic regime \cite{White} are determined from
the firing characteristics of a single self inhibited neuron (Figure \ref{Fig1}a). In the phasic regime
the network period depends on the synaptic decay constant ~\cite{White} and in the tonic regime the synaptic dynamics
weakly affect the network period. These two regimes of operation are clearly illustrated in
the plot of $\tau_{d}/T$, which is the ratio of synaptic decay constant to the firing frequency of the self
inhibited neuron versus $\tau_{d}$ in Figure \ref{Fig1}b for various values of $I_{In}$ and $g_{s}$.
These two regimes of network oscillations were observed earlier \cite{White}, with the synaptic model
for the inhibitory synapse, obeying the first order kinetic equation,
$\dot{S}(t)=F(V)(1-S(t))-S(t)/\tau_d$. Changing $\tau_d$ in this situation not only changes the decay time of
the synapse but also the rise time given by $\tau_{r}=\frac{\tau_{d}}{\tau_{d}F(V)+1}$ and the
saturation level of the synapse $S_{max}=\frac{\tau_{d}F(V)}{\tau_{d}F(V)+1}$. As a result a narrow region of higher harmonic phase locking was observed between the coupled interneurons in the phasic regime.
 We have therefore considered the synaptic model
presented above where we have control over the decay time for the
synapse independent of the rise time and the saturation level of
the synapse. Earlier work \cite{Chow} has shown that in presence
of mild heterogeneity, coherence in neuronal firing is observed
much more for phasic regime than in tonic regime. In the results
presented here we therefore
 consider the phasic regime of the network operation and study the effect of STDP in inhibitory synapses,
  in maintaining coherence in neuronal firing in presence of heterogeneity and noise. Details on the calculations
  for the tonic regime will appear elsewhere \cite{Talathi_Neurosci}

In all the subsequent calculations, we fix the parameters of the model in the phasic regime, ($I_{In}=2.5 \mu A/cm^{2}$,
 $g_{s}=g_{m}=0.1 mS/cm^{2}$, $\tau_{r}=1.1$ ms, $\tau_{d}=5.0$ ms) and study
the effect of the dynamic synapse in maintaining synchrony in presence of
heterogeneity and intrinsic noise. In figure \ref{Fig1}d, we plot the ratio of mean firing periods
 $\frac{<T_{B}>}{<T_{A}>}$ for the two coupled neurons A and B as function of heterogeneity in the external drive,
 defined as
$H=100\frac{I^{A}_{In}-I^{B}_{In}}{I^{A}_{In}}$ in absence of
noise i.e., $\eta=0$. As can be seen from Figure \ref{Fig1}d, (top
panel), the region of 1:1 locking, as function of heterogeneity is
much broader in the phasic regime of network operation as compared
to the tonic regime, where we set $I^{B}_{In}=5.0 \mu A/cm^{2}$.
In addition, higher order synchronization are also present in the
phasic regime, as a result coherence between the two neurons is
preserved more often in the phasic regime of the network
operation.

A spike timing dependent plasticity (STDP) rule for inhibitory synapses has been recently reported in
\cite{Haas} and it has the form
$\Delta g(\Delta t)=\frac{g_{0}}{g_{norm}}\alpha^{\beta}|\Delta t|\Delta t^{\beta -1}e^{-\alpha|\Delta t|}$,
 where $\Delta t=t_{post}-t_{pre}$. $t_{pre}$ is the time of presynaptic spike stimulation and $t_{post}$
 is the time of a spike generated by the postsynaptic neuron. $g_{0}$ is the scaling factor accounting for the
  amount of change in inhibitory
conductance induced by the synaptic plasticity rule and $g_{norm}=\beta e^{-\beta}$ is the normalizing constant.
 An empirical fit of the above function to the data gives,
$\alpha=1$ and $\beta=10$, giving a window of $\pm 20$ ms  over which the efficacy of synaptic plasticity is non
zero.  This implies that
in the physiologically important regime of gamma oscillations (25-80 Hz), STDP rule of inhibitory synapses
can play a significant role in modulating the firing dynamics of the neuronal network.

\begin{figure*}
\includegraphics[width=5.0in,scale=1,angle=0]{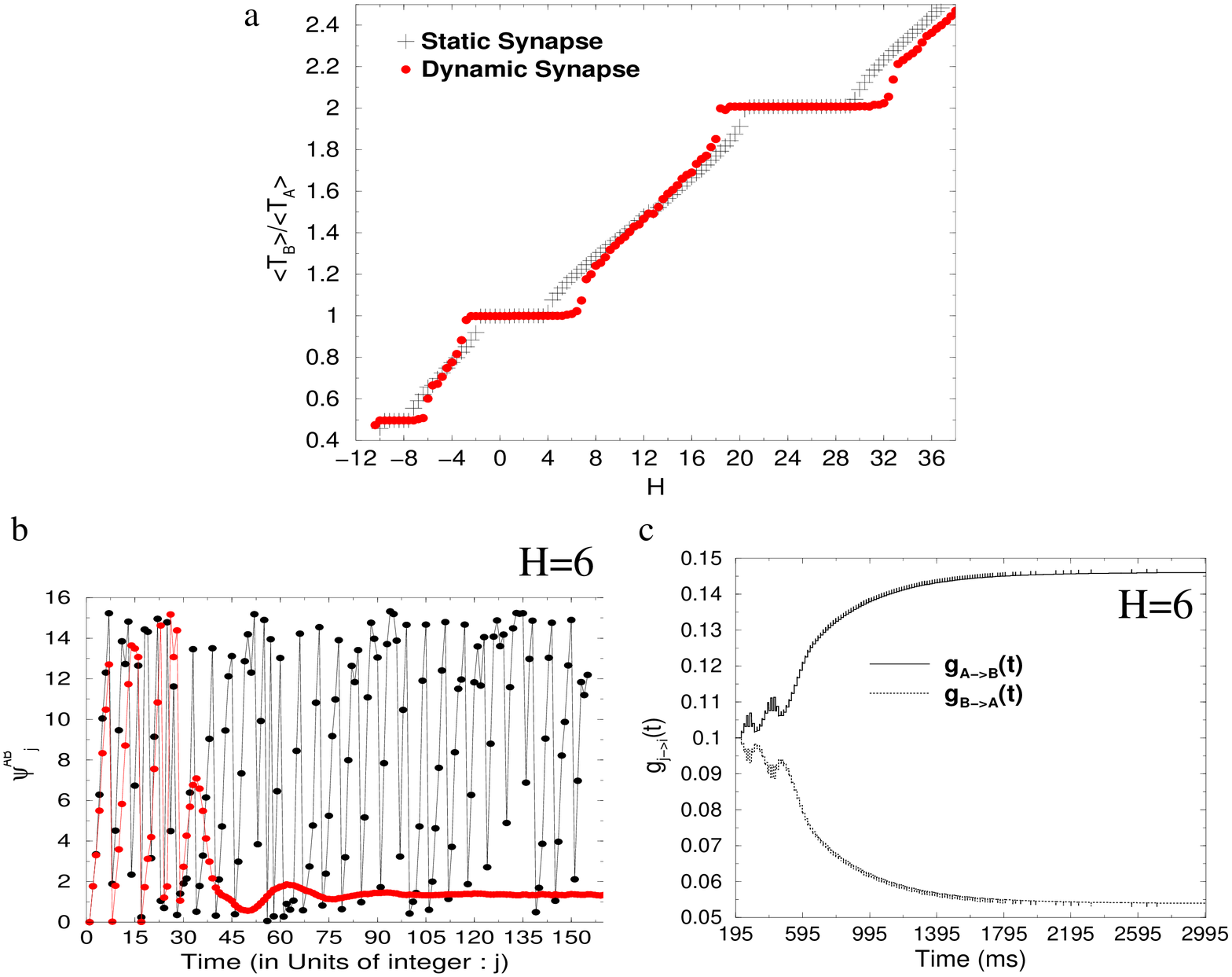}
\caption{(a) Ratio of firing periods of the two coupled
interneurons, in the static case, when the inhibitory synaptic
strength is constant and dynamic case when the inhibitory synaptic
strength is modulated by STDP rule, is plotted as function in
heterogeneity H. (b) The evolution of phase difference $\psi^{AB}$
is plotted in the static and dynamic case, when the heterogeneity
is set at 6 \%. (c) The evolution of the inhibitory synaptic
strength between the two coupled interneurons, A and B is plotted
as function of time. \label{Fig2}}
\end{figure*}

We now consider two situations in the phasic regime: when the strength of mutual inhibition is static:
 $g_{A\to B}=g_{B\to A}=g_{m}$ and when the mutual inhibition strength is dynamic and governed by observed
 synaptic plasticity rule, i.e., $g_{A\to B}=g_{m}(1+\frac{1}{g_{m}}\tilde{g}(t))$ and
$g_{B\to A}=g_{m}(1-\frac{1}{g_{m}}\tilde{g}(t))$. In order to take into account the effect of multiple spike
pairs, we follow \cite{Froemke} and define
$\tilde{g}(t)=\sum_{j}\sum_{i}\Delta g(\Delta t_{ij}) \epsilon^{A}_{i}\epsilon^{B}_{j}$, where
$\Delta t_{ij}=t^{B}_{j}-t^{A}_{i}$
is the difference in spike times of neurons A and B respectively. $\epsilon^{A,B}_{k}$ gives the efficacy of
spike in A and B and is defined as $\epsilon^{K}_{i}=e^{-(t^{K}_{i}-t^{K}_{i-1})/\tau_{e}}$.
We take $\tau_{e}\approx 55$ms,
an average of the efficacy values given in \cite{Froemke} as experimental results on contribution of multiple spike
pairing to inhibitory synaptic plasticity are as yet unknown. In Figure \ref{Fig2}a we plot the ratio
$<T_{B}>/<T_{A}>$ as function of heterogeneity H in the static and dynamic case.

As shown in Figure \ref{Fig2}a there is considerable increase in 1:1, 1:2 and 2:1 synchronization windows
mediated by the dynamic synapse in the phasic regime. This implies an increased probability
of observing coherence
in the firing pattern of the mutually coupled
interneurons even in presence of mild heterogeneity as has been reported in many in vivo experimental data.
 In Figure \ref{Fig2}b ,we show the evolution of phase difference
$\psi^{AB}_{i}=\left(|t^{B}_{i}-t^{A}_{i}|\mod <T_{A}>\right )$  in the static and dynamic case,
with heterogeneity of 6 \%. We see that in the static case the phase difference grows linearly
modulo $<T_{A}>$, representing situation of asynchronous firing. However in the dynamic case, after the initial
transient is over, the phase difference saturates to a fixed value, (1.4 in this example), representing stable 1:1
locking between the two mutually coupled neurons. In Figure \ref{Fig2}c we plot the evolution of the
synaptic strength as function of time. We see that the STDP rule results in modulating the synaptic strength
so as to phase lock the two mutually coupled neurons.

\begin{figure}[ht!]
\includegraphics[width=3.0in,scale=1,angle=0]{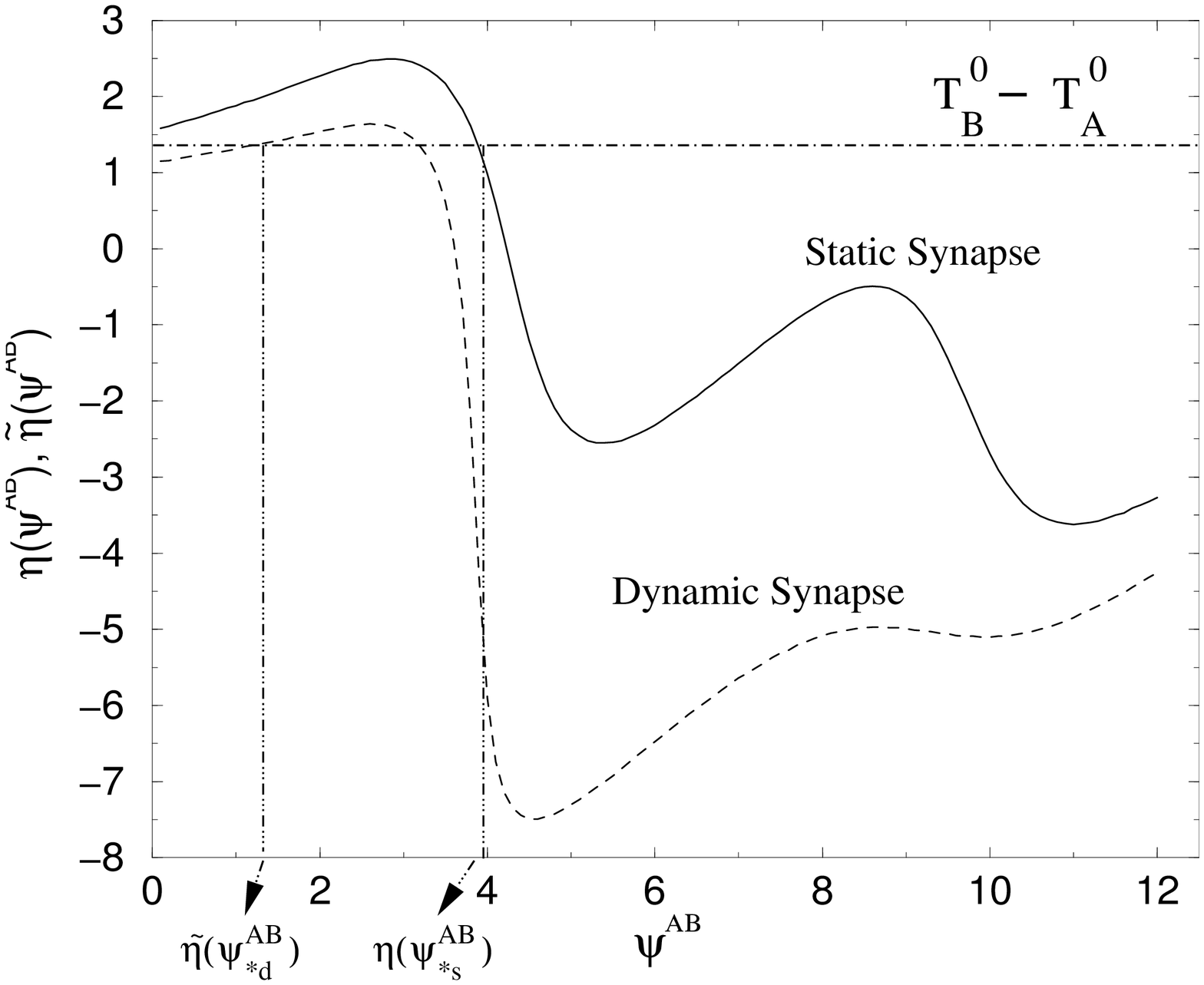}
\caption{Plot of the difference in phase response of the two coupled neurons in the static $\eta^{AB}$ and the dynamic
case, $\tilde{\eta}^{AB}$ is plotted as function of the phase difference. Stability analysis on $\eta^{AB}$
and $\tilde{\eta}^{AB}$ , determines whether the two coupled neurons will phase lock with each other. \label{Fig3}}
\end{figure}

In order to understand the dynamics of phase locking under
heterogeneity in the presence of dynamic synapse, we consider the
situation initially in the static case. Let the two coupled
neurons fire heterogeneously with intrinsic firing period
$T^{0}_{A}$ and $T^{0}_{B}$ when they are uncoupled. In the case
of 6 \% heterogeneity, we have $I^{A}_{In}>I^{B}_{In}$ so that
$T^{0}_{A}< T^{0}_{B}$. Let $\phi(\epsilon)$ be the phase response
curve of the neuron model. It is known that for type I neuron
models, the phase response curve is positive \cite{Ermentrout},
 and as a result every time a spike arrives, the phase of subsequent spike is delayed.  In the
 situation considered, with initial phase difference between the two neurons set to zero, the spike times of
 individual firing neurons can be written as,
  $t^{K}_{j}=t^{K}_{j-1}+T^{0}_{K}+\phi^{K}(\epsilon^{K}_{j-1})$, (K=A,B) with
 $\epsilon^{K}_{j-1} >0$,
 where $\epsilon^{A}_{j}=t^{B}_{j}-t^{A}_{j}$   and $\epsilon^{B}_{j}=t^{A}_{j+1}-t^{B}_{j}$. $t^{K}_{j}$ is the time
 of $j^{th}$ spike of neuron K. The map evolving the
 phase difference $\psi^{AB}$  is then,
\begin{eqnarray}
\psi^{AB}_{j}&=&(T^{0}_{B}-T^{0}_{A})+\psi^{AB}_{j-1}+\phi^{B}(t^{A}_{j}-t^{B}_{j-1}) \nonumber \\
&-&\phi^{A}(t^{B}_{j-1}-t^{A}_{j-1}) \nonumber \\
&\approx&(T^{0}_{B}-T^{0}_{A})+\psi^{AB}_{j-1}- \eta^{AB}(\psi^{AB}_{j-1})
\end{eqnarray}
 where $\eta^{AB}(x)=\phi^{A}(x)-\phi^{B}(<T_{A}>-x)$.
The fixed point of the map is then given by
$$\eta^{AB}(\psi^{AB}_{j})=T^{0}_{B}-T^{0}_{A} $$
Numerical solution to above equation gives the fixed points, at $\psi^{AB}_{*s}\approx 3.85$ as can be seen in Figure
\ref{Fig3}, where we plot $\eta(\psi^{AB})$ versus $\psi^{AB}$.
Stability of the fixed point would require
$|1-\frac{d\eta^{AB}(x)}{dx}|_{x=\psi^{AB}_{*s}}<1$ $\Rightarrow$
$0< \frac{d\eta^{AB}(x)}{dx}|_{x=\psi^{AB}_{*s}}<2$

For the set of parameters considered however, the numerical value obtained for
$\frac{d\eta^{AB}(x)}{dx}|_{x=\psi^{AB}_{*s}}=-2.5$ for the fixed point.

Thus we see that when the synapse is static the phase difference is unstable and the two coupled
interneurons fire asynchronously in the
situation of
6\% heterogeneity. Now consider the situation in the dynamic synapse case.
Again setting the initial phase difference zero, in the presence of dynamic synapse with STDP, we have
$t^{K}_{j}=t^{K}_{j-1}+T^{0}_{K}+\Hat{\phi}^{K}(\epsilon^{K}_{j-1},\tilde{g}_{j-1}(t))$, where phase shift given by
the phase response curve also depends on the dynamics of the synaptic coupling strength governed by the STDP
rule.
The map function for evolution of the phase shift $\psi^{AB}$ is then obtained as,
\begin{eqnarray}\psi^{AB}_{j}\approx(T^{0}_{B}-T^{0}_{A})+\psi^{AB}_{j-1}-\tilde{\eta}^{AB}(\psi^{AB}_{j-1})\end{eqnarray}
 where $\tilde{\eta}^{AB}(x)=\Hat{\phi}^{A}(x)-\Hat{\phi}^{B}(<T_{A}>-x)$.
 The fixed point of the map is then given by
 $\tilde{\eta}^{AB}(x)=T^{0}_{B}-T^{0}_{A}$.
As can be seen from Figure 3b, where we plot $\tilde{\eta}^{AB}(\psi^{AB})$, in the steady state ($t\to
\infty$), when $g_{B\to
A}=.146$ and $g_{B\to A}=0.054$, we obtain, $\psi^{AB}_{*d}\approx1.4$ and $\psi^{AB}_{*d}\approx 3.3$.
Stability of the fixed point given by
$0<\frac{d\tilde{\eta}^{AB}(x)}{dx}|_{x=\psi^{AB}_{*}}<2$, implies $\psi^{AB}_{*d}\approx1.4$ is stable as can
be also seen from the time evolution of phase in Figure \ref{Fig2}b.
Thus we see that STDP of inhibitory synapse, modulates the phase response curve such that the network locks into
synchrony even under mild heterogeneity.
\begin{figure}[ht!]
\includegraphics[width=2.0in,scale=1,angle=-90]{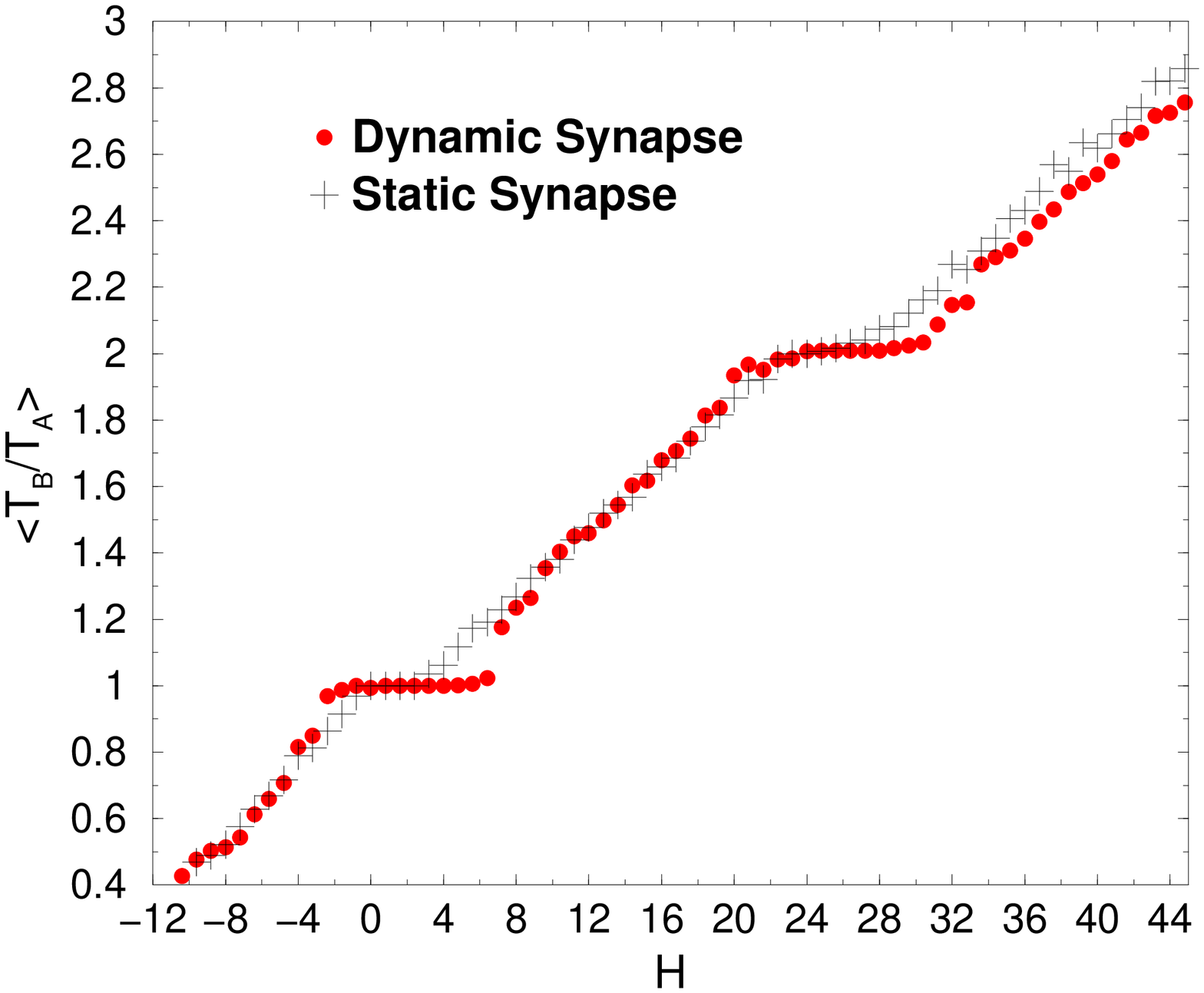}
\caption{Ratio of firing periods $<T_{B}>/<T_{A}>$ is plotted as function of heterogeneity H, in presence of
synaptic gaussian noise with amplitude a=0.1 \label{Fig4}}
\end{figure}
In Figure \ref{Fig4} we present results on synchrony in presence of noise. We set the noise amplitude
$\eta=0.1$ in equation 1.
For mild noise, STDP of inhibitory synapse, is able to maintain synchrony between the two
coupled interneurons under conditions of mild heterogeneity in the drive.

We have also tested the dynamics of the network in the tonic regime. STDP of inhibitory synapse, also
 significantly increases the window of synchronous oscillations by the same
mechanism \cite{Talathi_Neurosci}.

It has been suggested in \cite{Haas} that plasticity of inhibitory synapses may play an important role in
balancing the effect of excitatory synapse preventing run away behavior typically observed in
epileptogenesis. In this work we present an important function for STDP in inhibitory synapse in maintaining
synchrony in networks of coupled interneurons, under biologically realistic situation of mild heterogeneity
and noise.

{\bf Acknowledgements}\\
This work was partially funded by a grant from the National
Science Foundation, NSF PHY0097134. SST was partially supported by
the NSF sponsored Center for Theoretical Biological Physics at
UCSD. We would like to thank Henry Abarbanel, Julie Haas, Thomas
Nowotny and Johnathan Driscoll for instructive feedbacks that
improved this work significantly.

\end{document}